# Privacy Preserving Personal Assistant with On-Device Diarization and Spoken Dialogue System for Home and Beyond

**Gérard Chollet[1,2], Hugues Sansen[1,3], Yannis Tevissen[1,4], Jérôme Boudy[1], Mossaab Hariz[1], Christophe Lohr[5] and Fathy Yassa[2]**

[1]SAMOVAR, Télécom-SudParis, Institut Polytechnique de Paris, Evry 91011, France
[2]Speech Morphing Inc, San Jose, California 93128, USA
[3]Shankaa, Thiverval-Grignon,78850, France
[4]Newsbridge, Boulogne-Billancourt 92100, France
[5]IMT-Atlantique, Lab-STICC UMR 6285 CNRS, Brest, France

## Abstract

In the age of personal voice assistants, the question of privacy arises. These digital companions often lack memory of past interactions, while relying heavily on the internet for speech processing, raising privacy concerns. Modern smartphones now enable on-device speech processing, making cloud-based solutions unnecessary. Personal assistants for the elderly should excel at memory recall, especially in medical examinations. The e-ViTA project developed a versatile conversational application with local processing and speaker recognition. This paper highlights the importance of speaker diarization enriched with sensor data fusion for contextualized conversation preservation.
The use cases applied to the e-VITA project have shown that truly personalized dialogue is pivotal for individual voice assistants. Secure local processing and sensor data fusion ensure virtual companions meet individual user needs without compromising privacy or data security.

## INTRODUCTION

Since the availability of Siri on iPhones, a large number of "personal" vocal assistants are proposed on smart phones, smart speakers, smart watches, … But, are these truly "personal"? They all need an internet connection to capture your speech utterances in order to process them on the cloud. What are the cloud operators doing with these data? Even if it is true that you have given your informed consent to interact with these applications, there is still a risk that your speech data could be used for unwanted purposes like model training or exploitation of private information for targeted advertisement. Furthermore, none of these applications keep memory of previous interactions. They do not know who you are! In fact, they are quite "in-personal". But, with the availability of fast co-processors (GPUs and TPUs) on smartphones, there is no need to perform speech recognition and understanding on the cloud. The entire spoken dialog system can be embedded (Chollet et al. 2022). In the framework of the e-ViTA project (https://www.e-vita.coach/), we have developed such an application which does not need an internet connection to be operational.

Personal assistants for the elderly must bring functionalities that may not be useful for the digital natives. In particular, they have to be memory helpers for what has been said during a conversation. Among the identified use cases, the medical examination can be seen as the most obvious. By keeping track of the exchanges during the visit through speaker diarization, a personal assistant can build a context that can be interrogated later either by the person or by a caregiver, either at home or in nursing homes. Such a use case requires a real autonomy of a personal assistant, usually with no connection to the Internet, while keeping the privacy of the visit.

In the framework of the e-ViTA project, we have developed a spoken conversational application which could be used locally on android and IOS smartphones without an internet connection. It keeps track of previous interactions, recognizes who is speaking, performs automatic speech recognition, spoken language understanding, dialogue management and speech synthesis on the smartphone, searches the web if necessary, after anonymization of the requests, answers telephone calls, reads your emails, SMS, messages, prepares texts under your dictation, coaches your activities, behaves like a companion/butler and interacts with TalkMondo (https://talkmondo.com/) to facilitate inter-lingual communications.

Being perceived as less intrusive than facial recognition, vocal recognition/differentiation of the speaker requires less complex hardware since the device doesn't have to turn a camera towards a speaker. With the addition of a camera, which most smartphones have, the latter technique would need a beam-forming technique for spatial location of the speaker before orienting the camera, increasing the cost of a device and by extension of the maintenance.

In this paper, we emphasize the necessity of speaker diarization that allows the system to keep track of the user's conversations in a privacy preserving scheme. This technology, when deployed on an embedded device can also help track the elderly person's loneliness and give essential contextual information when a risk is detected by other sensors in a smart home for example.

## RELATED WORKS

There is a large number of vocal personal assistants available on smartphones, smart speakers and smart watches. They include Siri, Google Now and Google Assistant, Cortana, Alexa, Bixby, Elsa, SoundHound, Sherpa, Extreme, HeadUp, Jarvis, Voice Search, Voice Access, Android Auto, DataBot, Robin, Lyra, Otter, Fireflies, Murf, Socratic, Youper, Melissa, Api, Wit, Your Phone Companion, Friday, Tolkie, Vani, Vision, Bestee, 24me, Assistant, Dragon, EasilyDo, Indigo, Luiza, Reata and possibly many others,... Most of them need an internet connection to capture your speech signal and process it on the cloud. Shoebox (https://en.wikipedia.org/wiki/IBM_Shoebox) was an early exception by IBM in 1961 but it could only recognize the ten digits and 16 key words. Personal Digital Assistants (Psion, IBM Simon, Palm, Pocket PC,...) were popular in the 90s but they could not offer a vocal interface. Mycroft (https://mycroft.ai/) has been the world's first open source voice assistant which could be embedded on a desktop, in an automobile, on a Raspberry Pi,... and operates without connections. Snips (Coucke, A et al. 2018) planned to embed spoken language understanding on smartphones but was purchased by Sonos to be embedded in a smart speaker. It is only now that smartphones integrate co-processors which are capable of real time speech recognition, understanding and synthesis.

## USE CASES

While on, an individual vocal assistant (Iva) could start a dialogue with an unknown person and reveal information about the owner. The following use cases make the hypothesis that the persons around the owner accept being recorded. In this text we describe a simple day of interaction with Iva, with each phase being a use case of its own. When the smartwatch and the sensors located in the room detect an activity, early in the morning, Iva welcomes the owner by wishing a good day, recalling an appointment with the medical doctor at 10 AM and giving the weather forecast for the day. Arriving at the doctor's, Iva recognizes a new speaker and asks the speaker's name and activity. Until the end of the consultation, Iva doesn't intervene unless she is asked to. She just transcripts the dialogue between both persons. At a certain point, the doctor can ask Iva what the heart rate of the owner was when she woke up. Back home, the owner can ask an explanation related to the consultation, in less technical terms. Enters a caregiver. The caregiver asks Iva to summarize the consultation. Iva recognizes the caregiver and adapts the summary according to the needs of the caregiver. Would it be a housekeeper, Iva would respond whether the owner is in good health, if ill, contagious or not. Later in the day, the owner asks a question about history. Iva makes a Web search and answers the question. Later in the day, Iva interrogates the owner about her life, an endless source of dialogue. Days after days, enough information is gathered to reconstitute the life of the owner and write her biography, her lifeline [López A. et al., 2017]. This "LifeLine" feature is a short term/long term memory stimulation for the owner and a posthume gift for her family. Moreover, Iva is never bothered by being told the same story for the $n^{th}$ time, making Iva the best confidant one could dream of.

For John, this is another case. John is in his late sixties and still in good shape, at least he thinks he is, except that he has had to visit the bathroom many times each night for a few weeks. Besides having poor sleep, he is ashamed of the situation. He reassures himself into believing it will naturally improve. The sensors installed in his house as well as his smartwatch detect his nightly activity and report it to the e-VITA application embedded on his phone, not on the Internet. Because it is an obvious health alarm, for a few days, e-VITA has been recommending that he should consult a medical doctor. At the moment, this is still a secret between him and his phone. A secret to which no software vendors have access.

Some use cases have been left aside for the moment, such as how to avoid the service to engage in a dialogue with the TV or with another vocal assistant. The latter case is of course weird, but can also be used as a way to train and improve the service. Finally, the benefits of being able to be used without having to connect to a cloud server through the Internet has its benefits on its own as the service can be used out of any connection, like during outage or when no reliable connection to a 5G network is available.

Besides the usage, a designer must also take into account the operating cost of a service and its deployment. By integrating most of the processing on a device that is ubiquitous, we get rid of the operation of a server and we can benefit from the deployment platforms used for those devices. This contributes to making the service affordable for the targeted users, with a simple subscription model.

## SYSTEM ARCHITECTURE

Benefiting from the processing power of modern smartphones, the architecture of the assistant embeds all the processing of the services inside an Android device (Pixel 6/7/8) (figure 1). To preserve the privacy of the dialogue exchanges, and to allow the usage in a non connected mode, all private interactions are performed on the device, with no exchange of private data with Cloud servers.

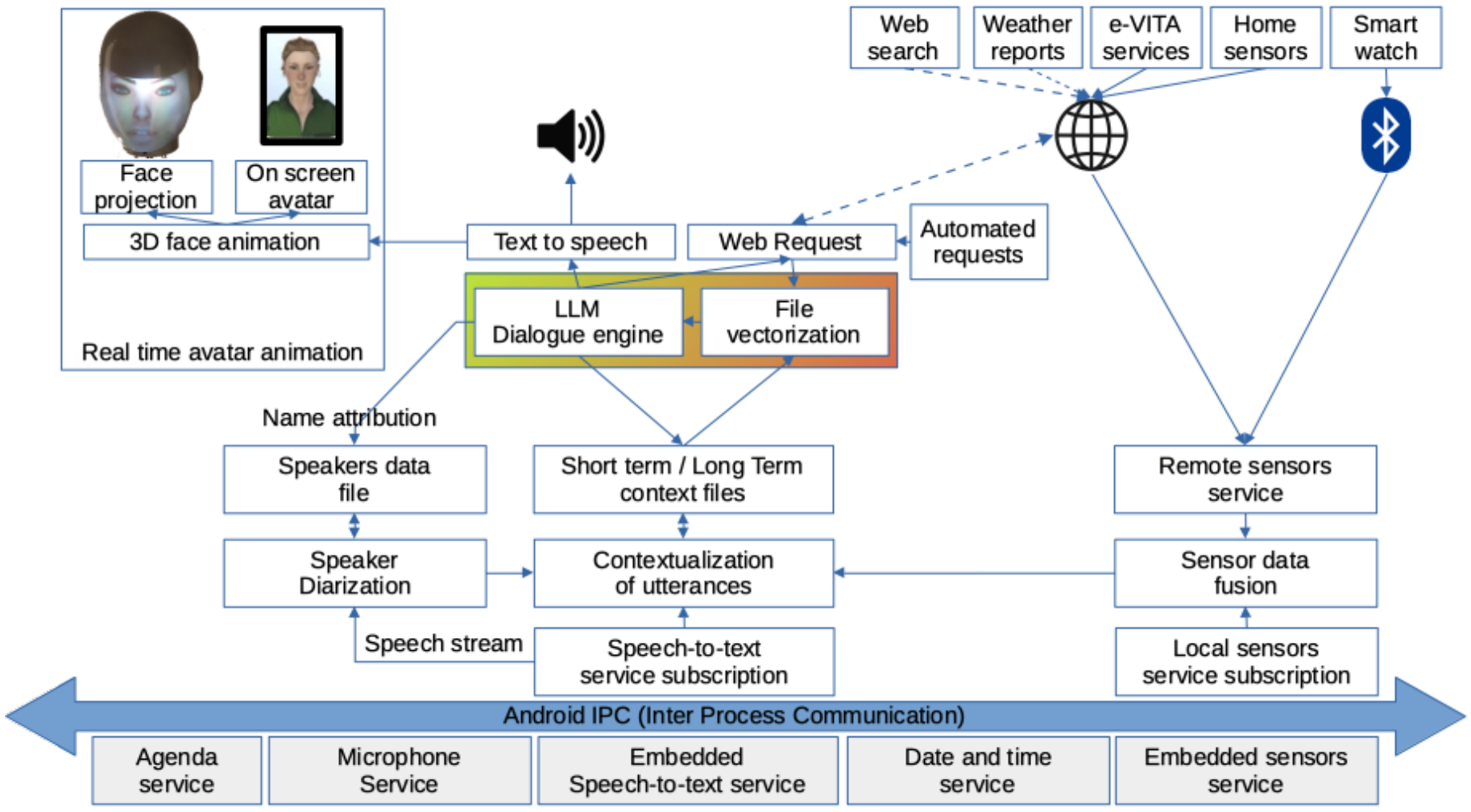

**Figure 1:** Ideal architecture blueprint of the Individual Voice Assistant on an Android phone (for clarity not all relationships are represented)

This architecture preludes the integration of talking faces initiated in the Empathic project [Torres M.I. et al, 2019] to animate a face in real time (figure 1 top left). The very same architecture will be used in the animation of the projected face of the robot "Roberta Ironside" [Sansen H. et al., 2016]. By embedding all the logic inside a mass market device, such as a high end smartphone, the architecture allows the conception of a physical companion/embedded avatar that will be affordable for a population with limited resources while facilitating the maintenance and the reliability.

Unfortunately, many Android security features prevent a simple realization of this architecture:

- Google speech recognizer doesn't return the audio stream of the speech being recognized, the callback is just optional.
- It is not possible to allocate the microphone simultaneously to the diarization and the speech recognizer. A solution that could have been a workaround to the previous point, but that is not possible for obvious security policies.

As a result, either the project will have to abandon the diarization, or wait for Google integrated diarization that has been demonstrated in 2019 [El Shafey L. et al., 2019]. The third option being the integration of a custom speech recognition, a project in itself that may be a last resort which is currently under investigation with partners.

## SENSORS DATA FUSION

Data fusion is the process by which data from different sources are combined to obtain new, higher-level knowledge. Thanks to the proposed architecture, data fusion can be managed both at cloud level and at the edge; by edge device, we mean here the smartphone that implements our system.

Firstly, the system can make web requests to a cloud server. As a server, the e-VITA platform collects data from the user's environment, performs data fusion processes and makes the results available to the user's services (Szczepaniak et al. 2023).

For example, e-VITA manages infrared motion sensors, door-opening sensors, indoor temperature and climate, and so on. The data minimization and data fusion component integrated in the platform provides statistics on the amount of time the user spends in each room (and deduces a level of sedentarization), but also allows proceeding algorithms for Human Activity Recognition (HAR) which produces labels with timing such as "cooking" "eating" "resting" "toileting" etc.

Those raw data and computed labels can be retrieved by the smartphone, after authentication, as for any other web queries. Since it is not mandatory for user services to get at real-time a report on these activities over time, it can be chosen to download this sporadically.

Secondly, the proposed system enables the smartphone itself to fuse data. The smartphone is a source of data (e.g. accelerometer, gyroscope and magnetometer), which can be analysed by itself to produce labels about the user's pose ("standing" "sitting" "lying" "walking"). The user's activity (at least time of the day) with the coaching application is also a source of data and knowledge.

Finally, the data fusion embedded on the smartphone can combine information from the web and itself. For example, it can deduce the preferred context (location, pose, time) when the user likes to chat with the coaching application.

## SPEAKER RECOGNITION AND DIARIZATION

As the elderly person interacts with the assistant on a daily basis, there will be many situations where this person won’t be alone and possibly not the only one speaking. Therefore, the system needs to be able to focus on its primary user voice. Such technologies exist and are already embedded in several home assistants such as the Google Home that possess an implementation of the Personal VAD technology (Ding et al. 2020).

Moreover, the system needs to understand the conversational context in which it is evolving. It needs to keep track of the conversations of its user to provide more relevant answers. For example, it should be able to summarize a conversation between the user and his doctor to help the former with his treatment or diet (Riad et al. 2022).

For this to happen in a privacy-preserving scheme, speaker diarization is directly implemented on the device. Performing speaker diarization is equivalent to answering the question of “who spoke, and when?”. By extracting speaker turns, the system can easily isolate one speaker’s instructions and discard the ambient speech if it has no interest for the assistant. Recent speaker diarization methods are based on several deep learning modules that segment the audio into small atomic segments, then create speaker embeddings for each segment and finally cluster them to find the different speaker relative identity.

## SPOKEN DIALOGUE SYSTEMS AND LLMS

In parallel with our developments, the e-ViTA partners have developed a more traditional approach based on a private cloud server to satisfy the needs of the robotic systems being experimented (Mc Tear et al. 2023). Several of these devices (Nao, Gatebox, Daruma, SanTO and Android robots) use cloud based interfaces by default. The smartphone and the tablet were chosen in addition to facilitate privacy preserving experimentations. The Google Pixel 6, 7 and 8 offer an embedded ASR running on TPU with higher performances than cloud based solutions.

The architecture embeds a LLM model on the phone inside Android code that reproduces some functionalities of Langchain (https://www.langchain.com/). Domain specific files are introduced and vectorized to constitute a permanent vector base. Temporary files, e.g. Web documents, are added in a temporary vector base as they are received. The vector bases used for the day are merged into a workspace vector base to serve as the workspace context. Additional events, diarization, sensor events, Web searches, are added to a temporary vector base that is merged to the workspace vector base as they arrive. Despite being powerful, the devices (Pixel 6/7/8) have a RAM of 8GB (12GB for the Pro version) while LLama2-7b (16GB) requires 6.52GB of RAM when quantized in 8 bits, a technique introduced in 2018 [Polino, A. et al., 2018]. In order to keep enough memory space for other applications, it is necessary to lower the quantization to 4 bits or less. Elias Frantar et al. have shown that their “*method can still provide reasonable accuracy in the extreme quantization regime, in which weights are quantized to 2-bit or even ternary quantization levels*" [Frantar E. et al, 2022].

Google’s TensorFlow-Lite, i.e. TensorFlow for mobiles, has been designed to work with 8byte quantized models and is supposed to use the TPU of Google Pixel phones. The Keras formatted GPT2 model that is used to answer questions, occupies 1.5GB once quantized. Even if GPT2 on TensorFlow-Lite is sufficient for a proof of concept, the TPU is not used and the response time, ranging from 15s to more than a minute on a Pixel 6 or a Pixel 7 is incompatible for a serious dialogue. As is, the solution cannot be proposed as a product. Thus we will have to wait for the availability of Gemini, an evolution of Bart [Gemini Team, 2023], in its mobile version: Gemini nano for the Pixel 8 Pro in Europe. This should happen when Google demonstrates the compatibility with the recent regulations of the European Union on AI. The drawbacks are a price tag around €1000 for the Pixel 8 Pro that will significantly increase a potential subscription fee for the service, and the uncertainty of the truly delivered functionalities.

As an alternative to TensorFlow-Lite, Llama CPP, an Open Source C++ implementation of Llama, designed to run on CPUs, is under investigation. This solution has to be integrated on Android in its 4 bit quantization version. Without the usage of the TPU, there is no hope to obtain a better response time, though, as tests on PCs show a response time around 1mn.

With context documents covering the targeted domain, and with a good tuning, the answers are sufficiently reliable for a concise dialogue, while larger models may be out of reach. To deal with the response time, a smaller model rephrases the questions, and the "large model" constructs the answers while the TTS reads the text of the rephrased question. Despite positive results, improvements are needed. In the project roadmap, Llama 2, of which the availability on Keras was announced last July, will be tested after quantization when the application is stable.

Because not all answers of the dialogue system are meant to be given to the user, they are filtered according to a marker. Sensor data are an example of events that do not require answers except when they are out of a given pattern. When an unregistered speaker is detected by the speaker diarization, the utterance is marked as such and triggers a question by the dialogue system to request the name of the speaker. Those that are marked as responses to the questions of a speaker are forwarded to a text-to-speech module. Since the native TTS service of the Android library doesn't support prosody annotations, Speech Morphing TTS is preferred. It is a commercial solution in which two of our authors are involved. A similar approach has been used in the Empathic project in conjunction with the talking faces [Torres M.I. et al, 2019].

## AUTOMATED ACTIONS

Some actions such as weather report retrieval, agenda for the day announcement, are automated so that they can be announced at the start of the application. Such actions run in a background process even while the application is not running. At the moment, we are still questioning whether the service should wake up when unusual events are detected by the sensors.

## DISCUSSIONS AND UPCOMING CHALLENGES

The Open Voice Network Association (https://openvoicenetwork.org/), W3C (https://www.w3.org/) and MPAI (https://mpai.community/) are developing recommendations and standards that we are contributing to. The need for a cloudless approach that would protect the user's privacy seems shared by other teams such as Apple's [Aas, C. et al., 2023].

The e-Vita project targets a population that may not feel the need for changing its habits. As always, the real difficulty is not applying a technology but to make it relevant and easy to use, especially when dealing with non digital natives. At the moment the installation of the embedded service involves complex operations that are acceptable in an evaluation phase. The personalization of the settings presents another source of difficulties for the user, even for people comfortable with the technology. To be successful we will need to adapt services to fit the needs and culture of a specific population as well as to make the application easy to install. For the settings, we can use the vocal interaction with the service, but even if this interaction is facilitated, the verbatim must also be easily understandable. The same way that people may like or dislike the announcements of their GPS navigation system, (in 200 meters, turn left!), a personal assistant must adapt its volubility to the user, a conundrum even in real life.

The start and stop of the application necessitates a physical action on the phone. In some cases, it would be interesting to start or stop it vocally or stop it when there is no vocal ambiance. This involves a background process that consumes energy when the system is idle. Some analog solutions, such as Aspinity's AML100, exist but are not integrated in smartphones. While possible on a robot as a plug-in device, they may not appear on smartphones before a while unless a smartphone OS provider includes it in its roadmap.

On the other hand, leaving some services to mobile OS providers presents the risk of leaving the management of privacy preservation to big companies as well as a potential risk of indoctrination of the users. A risk that an independent designer of a service may not be protected from either. Can we imagine the designer being trialed for having built a service that provided malicious advice? As for autonomous cars, the e-Vita project is not immune from an accident due to AI, despite the best intentions. Unfortunately, usages have shown that even restrictions applied to AI systems can be worked around by asking questions differently. Some issues are addressed by papers about voluntarily or involuntarily malicious usages of AI such as [Brundage M. et al. 2018], but the extreme complexity of data in current AI systems may bring hurdles that cannot be solved.

Technically, embedding a large language model on a device, as powerful as the recent high end models, is a challenge. It will necessitate fine tuning at every level of the different modules of the application. Some capacities available with the e-Vita cloud server will be degraded on its embedded counterpart.

It is not clear at the moment if, by distributing the processes on the client side, the global carbon footprint is reduced compared to the processing on a centralized server. The first option consumes energy only when in use while the latter is always on, even when no processes are active. On the other hand, storing energy on a battery is not environment-friendly, and, when used, mass market wireless charging doesn't have a good efficiency. A future evolution could be the monitoring of the battery consumption by the application even if this feature may not be perceived positively by the marketing.

## CONCLUSIONS

The computing power of smartphones allows for vocal interactions without an Internet connection. Automatic Speech Recognition, Spoken Language Understanding and Speech Synthesis can be embedded with improved performances. In particular, the latency is considerably reduced and the models can be adapted to the user.

These recently available technical capacities of smartphones allow service designers to fulfil usage constraints such as the guarantee of the encapsulation of private data on a device owned by the user. This guarantee is the primary goal of the embedding of the service on the client side. In this process, the diarization of the speaker plays a major role, as it allows the management of who can interact with the service and in some cases adapt the level of privacy with which the application can answer a question.

Affordability is a major concern for the project and the targeted users. By relocating the processing on the user device, the exploitation model is streamlined and the operating costs are reduced, making the service more affordable for the user. This raises a question about the greenness of such a service, an issue that all designers must keep in mind. Unfortunately, by distributing the energy consumption, it becomes hard to evaluate the real impact on the environment, and centralized systems are supposed to be more virtuous than distributed ones. On the other hand, a developer, who has to deal with the limited resources of the mobiles, is naturally more conscious of power consumption. Privacy preservation or greenness, this is the dilemma.

## ACKNOWLEDGEMENTS

The authors would like to acknowledge the financial support of the European Commission and the Japanese MIC through the H2020 e-ViTA project (https://www.e-vita.coach/) and the contributions of Florian Szczepaniak in the early stages of this project.